# Multicolor localization microscopy by deep learning

Eran Hershko[1,*], Lucien E. Weiss[2,*], Tomer Michaeli[1], Yoav Shechtman[2]
*Equal contribution
[1]Electrical Engineering Department, Technion, 32000 Haifa, Israel
[2]Biomedical Engineering Department, Technion, 32000 Haifa, Israel
Corresponding author: Yoav Shechtman, yoavsh@bm.technion.ac.il

**Abstract**

Deep learning has become an extremely effective tool for image classification and image restoration problems. Here, we apply deep learning to microscopy, and demonstrate how neural networks can exploit the chromatic dependence of the point-spread function to classify the colors of single emitters imaged on a grayscale camera. While existing single-molecule methods for spectral classification require additional optical elements in the emission path, *e.g.* spectral filters, prisms, or phase masks, our neural net correctly identifies static as well as mobile emitters with high efficiency using a standard, unmodified single-channel configuration. Furthermore, we demonstrate how deep learning can be used to design phase-modulating elements that, when implemented into the imaging path, result in further improved color differentiation between species.

## Introduction

Single-particle tracking and super-resolution fluorescence microscopy harness a high signal-to-background ratio to attain nanoscale spatial information exceeding the diffraction limit. Localization-based microscopy techniques ((F)PALM, STORM[1–3], and related methods[4–7]), can improve the resolution of an image by an order of magnitude relative to that attained in normal epifluorescence microscopy, routinely attaining $10 - 40\ nm$ spatial resolution. The key idea of localization microscopy is to find the likely underlying position of an emitter whose emission produces a diffraction-limited spot captured by a camera (e.g. the center of the point-spread function, PSF). It has been shown previously, that in addition to the position, other information can be extracted from the PSF as well, such as photophysical properties[8], molecular orientation[9], and directionality of motion[10].

Of particular importance for biological imaging is real-time, correlative information between multiple species in a sample[11] (e.g. proteins or organelles in cells and tissues). Typically, this is achieved by attaching spectrally-distinctive fluorophores to molecules of interest, thus necessitating multicolor imaging. Using an RGB camera is not practical for low-signal applications where photons are precious, such as single-particle tracking and single-molecule microscopy. The popular approaches for multicolor microscopy either divide the emission spectra between multiple cameras or regions on the same camera[12,13], or sequentially image one species at a time by switching spectral filters and/or light sources[11]. The former method requires precise registration between the channels[14], while the latter is limited to imaging quasi-static objects and is prone to artifacts caused by non-simultaneous, slower acquisitions, e.g. sample drift.

Recently, multicolor localization microscopy by PSF engineering has been demonstrated[15,16]. In this technique, the image that each point source creates on the camera, namely, the PSF, is modified to encode the color of the emitter. In other words – molecules emitting different colors produce images of different shapes. This modification is performed by adding a spectrally-sensitive, phase-modulating element, e.g. a liquid-crystal spatial light modulator (SLM) to the imaging path which is positioned in a plane conjugate to the back focal plane of the microscope objective[17]. The main advantage of PSF engineering is that it enables truly simultaneous imaging and tracking of multiple, colored emitters with no compromise in the field-of-view (FOV). However, the design of a multicolor PSF that optimally balances between emitter detectability, which requires the PSF to be small, and color-classification, which requires the PSF to vary significantly as a function of wavelength, is still an open question. Furthermore, the additional optical elements required for PSF engineering (or any existing simultaneous-multicolor imaging approach) add complexity to the microscope, limiting the applicability for general use – clearly, achieving simultaneous multicolor localization over a large FOV using a standard microscope would be highly advantageous.

In recent years, deep learning has shown great success in a variety of tasks[18], including designing optical systems[19–21] and interpreting single-molecule data to produce super-resolution images[22,23]. The multi-layer architecture of neural nets allows for extraction of complex features from data, while distilling the desired information from an input. Neural nets are capable of learning to recognize subtle features, even under adverse imaging conditions, making the technique well suited for the problem of species differentiation in localization microscopy where the PSF differs only slightly between different wavelengths.

Here, we present two fundamental contributions by applying deep learning to microscopy. First, we experimentally demonstrate an algorithm for determining an emitter's color using a standard fluorescence microscope equipped with a grayscale camera with no additional hardware modification (Fig. 1). This is enabled by the fact that the PSF of any optical system is dependent on the wavelength, even without PSF engineering. Second, we develop and experimentally demonstrate an additional neural net that algorithmically optimizes a color-encoding PSF using phase modulation, for maximal color-distinguishability.

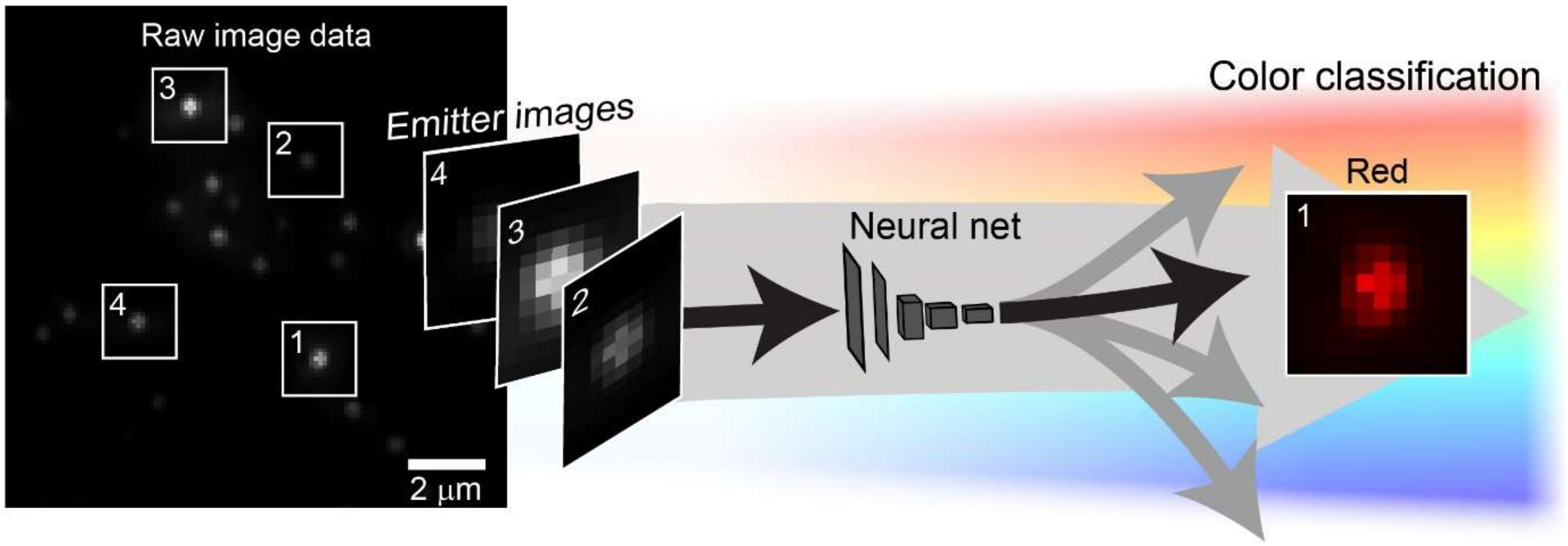

**Fig. 1** Color classification with neural nets. Patches are extracted from a grayscale image and classified by a neural net. Red and green quantum dots shown here.

## Results

The PSF of any imaging system depends on the emitter's wavelength, due to diffraction and possible chromatic aberrations. To test whether a neural net could discriminate between two types of emitters, we prepared a thin sample containing green and red quantum dots (Qdots) with emission peaks at 565 and 705 nm, and imaged it using an epifluorescence microscope (Fig. 2a). For each field of view (FOV), three image sets were recorded: first, a grayscale image containing all of the Qdots (Fig. 2b); second, an image with a spectral long-pass filter added so that only the red Qdots were visible (Fig. 2c, red channel); and lastly, a bandpass filter so that only the green Qdots were visible (Fig. 2c, green channel). A deep convolutional neural net was trained using twenty such FOVs, containing ~400 Qdots per field (training details are described in methods, Fig. S1). The net was then used to classify Qdots in a new FOV (Fig. 2d). Due to blinking, only the Qdots visible in both the Ground truth and Raw data images were classified. We attempted to identify the distinguishing characteristics of the two PSFs, but could not find a clear separable difference between the red and green PSFs in terms of emitter brightness (Fig. 2e) or the parameters of astigmatic 2D Gaussian fitting (Fig. 2f). Using subpixel shifting to align patches in the image, we computed the mean PSF for the red and green Qdots, then compared the performance of a matched filter to that of the net (Fig. 2h). The matched filter only performed slightly better than a random assignment (i.e. ~55%), while the trained net correctly predicted 96.4% ± 1.2% (N= 2491).

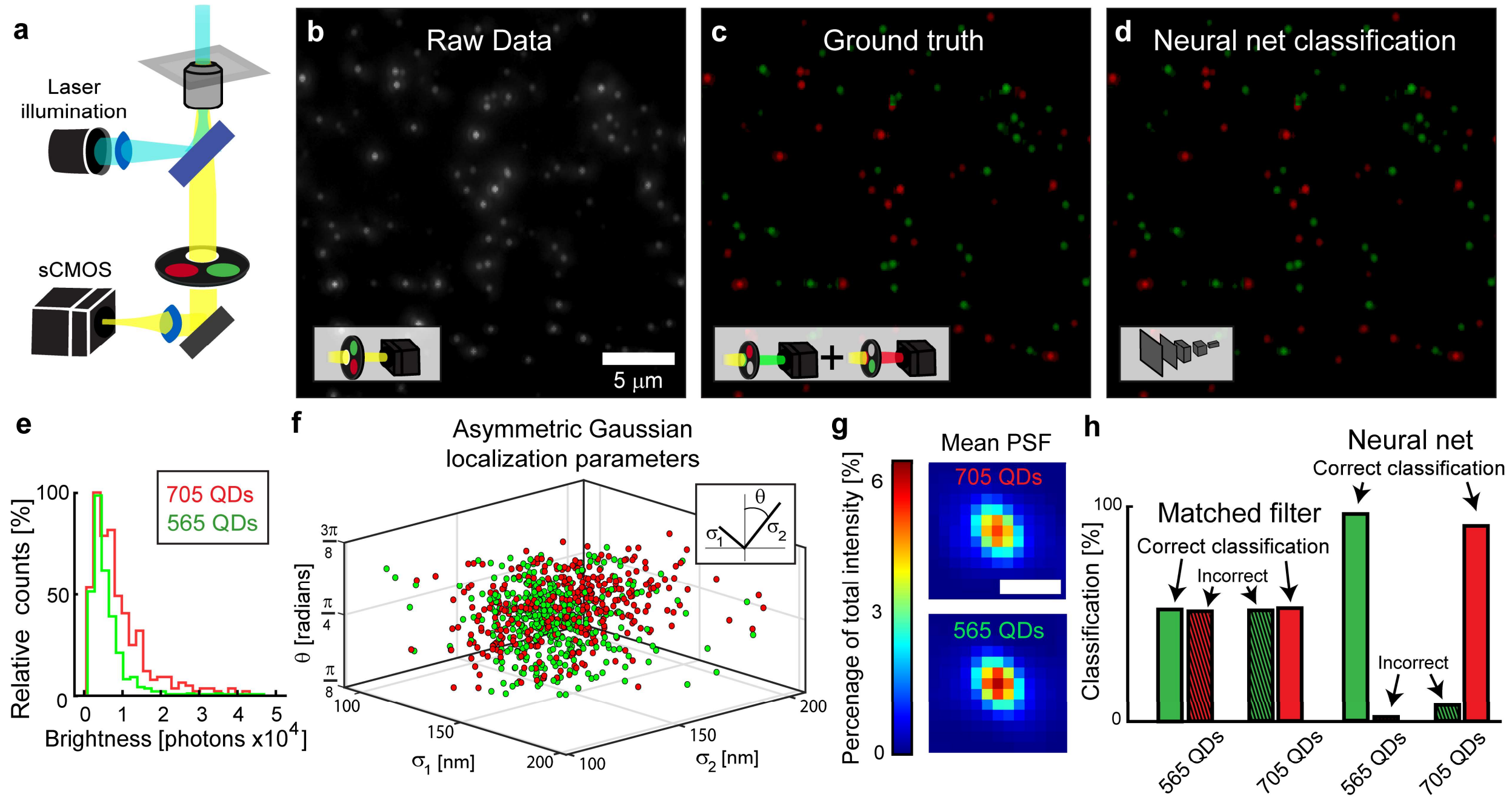

**Fig. 2** Qdot color determination using neural nets. **a** An epi-illumination microscope was used to examine Qdots on a glass coverslip. **b** A grayscale image of red and green Qdots. **c** A color image of the same sample obtained by imaging with two spectral filters. **d** The color classified grayscale image obtained from the neural net. **e** A histogram of the signal photons of the two quantum dots. **f** A 3D scatter plot of the red and green Qdots showing the fitted parameters from an astigmatic Gaussian with two shape parameters ($\sigma_1$ and $\sigma_2$) and a variable angle ($\theta$). **g** Average PSFs for red and green Qdots. **h** Classification percentage for emitters by correlation with the average PSFs and by neural net classification.

Next, we checked how well a neural net can handle a more challenging problem: determining the identity of moving particles, where the PSF in each frame is convolved with the trajectory within the exposure time, i.e. motion blur. In brief, samples for tracking experiments were prepared by squeezing a 10 uL droplet containing two types of fluorescent, sub-diffraction-sized beads between two glass coverslips (red beads: 645/665 nm absorption/emission; and green beads: 505/515 nm absorption/emission). The sample was then imaged with a 20X Air objective, NA 0.75, producing a quasi-2D diffusion chamber ~2 μm in height, which was similar to the depth-of-field of the imaging system (Fig. 3a).

To gather training data for the net, three types of movies were recorded sequentially: images containing both colors of beads (Fig. 3d), as well as red-filtered and green-filtered images; this cycle was repeated several times to ensure that the ground truth was known (shown as a maximum intensity projection Fig. 3e). Unlike the immobile Qdot data, the diffusing beads yield a more time-varying data set that can be treated as many independent measurements of the PSF. The net was trained using sequential frames of a centered ROI around each bead (Fig. 3b) obtained in 17 separate movies, totaling 748 red and 684 green beads. Unlike the single-frame determination used for the static Qdots, a single image alone was typically not sufficient to determine the color (Fig. 3c). Instead, the net performed best when the input was 100 successive frames per emitter. To evaluate the performance, the net was used to determine the color of 579 beads imaged in 5 movies (93.8%± 3% were correctly classified). Not all beads in the sample were mobile, and the net predicted the identity of static and mobile particles with different, albeit high efficiencies (96%± 2% of the 393 mobile particles, 89.2%± 3.6% of the 186 static particles).

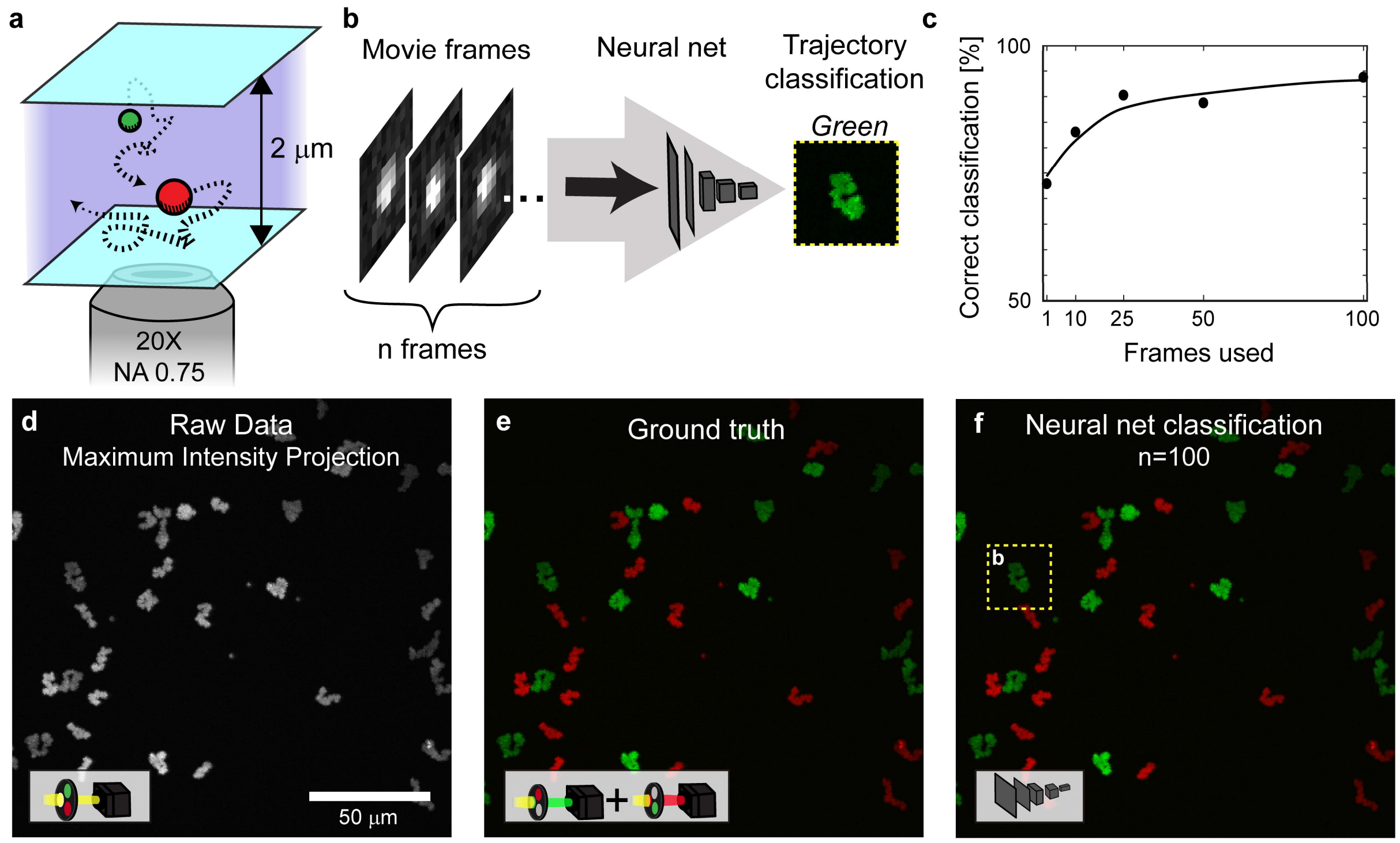

**Fig. 3** Color determination of moving microspheres. **a** Illustration of the diffusion chamber. **b** Schematic of the neural net classifying sequential groups of frames belonging to the same emitter as red or green. **c** The performance of the net as a function of the number of frames used for classification. **d** A maximum-intensity projection of diffusing beads. **e** Colorized intensity projection from red and green filtered images. **f** Neural net classification based on n=100 sequential frames per bead.

Qdots and fluorescent beads are bright emitters compared to single molecules. To demonstrate the potential of our approach to single-molecule localization microscopy, we both simulated single-molecule blinking data (Fig. 4a-c) and imaged fluorescently-labeled HeLa cells (Fig. 4d-f). In simulated data, octagonal structures, containing either Alexa 568 or Alexa 647, were randomly placed in a FOV using the TestSTORM toolbox[24]. Grayscale images for analysis were created by adding Red-only and Green-only frames together (Fig. 4b). The net-correctly classified 90% of emitters (Additional simulations in Fig. S9).

In experimentally obtained data, cells were fixed and then the microtubules and mitochondria were fluorescently labeled with Alexa-647 and Alexa-555 (Fig. 4d). Immediately before image acquisition, the sample media was replaced with imaging buffer for blinking single molecules as described previously[15]. Net training was done using regions of the image containing only one species, as identified by diffraction-limited images, or in single-antibody labeled samples. Only molecules with a high SNR (>8) were used for training. A super-resolved image is reconstructed by creating a 2D histogram of the localized positions (Fig. 4e), where each bin was colored using the net's classification score of nearby emitters (Fig. 4f). Beyond the successful color-determination in the zoomed-in region of Fig. 4f, the current limitations of our method are also visible. In highly-challenging conditions (e.g. the top part of the cell Fig. 4d-f), the color-differentiation did not match the diffraction-limited image. This may be due to the higher density of emitters, increased sample thickness, and non-uniform imaging conditions across the FOV.

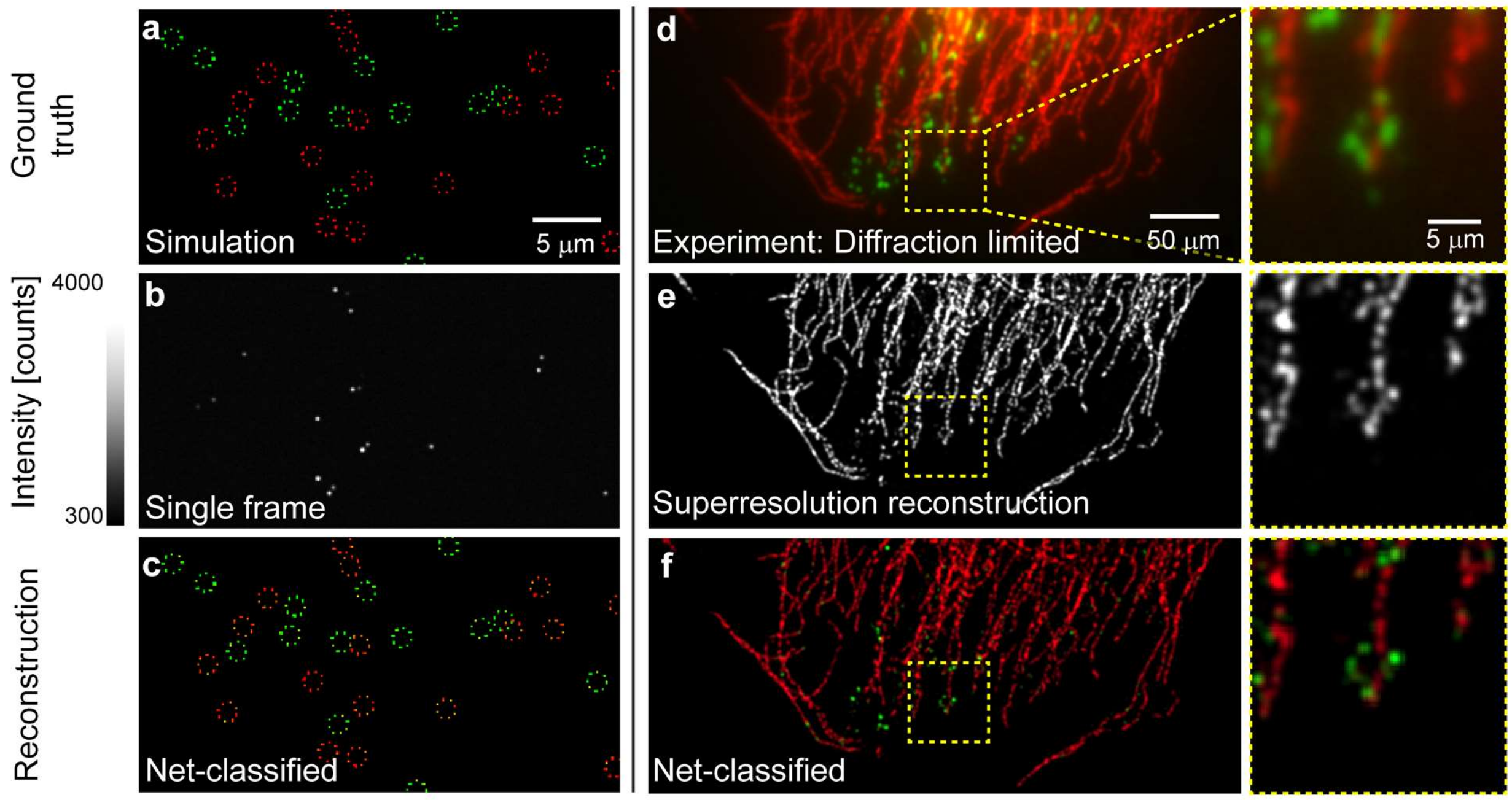

**Fig. 4** Single-molecule reconstruction based on net classification. **a** Ground truth of simulated red and green structures. **b** An example single frame containing red & green emitters of the region shown in a. **c** Classification by the net. **d** Combined spectrally-filtered, diffraction-limited image. **e** Super-resolution reconstruction. **f** Reconstructed image, with pixels colored according to the net-classifications of individual localizations.

Until this point, we have only utilized the standard PSF and have not made explicit use of any strongly wavelength-dependent optical elements, such as prisms or liquid-crystal SLMs (Fig. 5a). An SLM in the imaging path can add spectrally dependent aberrations to the PSF that could help distinguish between different colors[15,25], but what is the best way to do so? Namely, what is the optimal trade-off between high-localizability (requiring a small PSF footprint) and efficient color-distinguishability (requiring spatially different PSFs for different colors)? Previous work on optimal PSF design for 3D imaging used Fisher information as a design metric[26,27]. Here, we solve the PSF design challenge using deep learning.

To engineer an optimal PSF for color determination, we simulated emitters to train a physical net, which was used to determine the PSF-shaping, while simultaneously training a reconstruction net to recover the colors and positions (Fig. 5b). At each iteration, an SLM voltage pattern is generated by the physical net, creating a wavelength-dependent PSF. Simulated red and green point sources at random positions are fed into this net to produce a camera frame in which each point source appears as the PSF corresponding to its color. Next, these images are fed to the reconstruction net to produce a colored localization map on a grid which is 4X finer than the original image. The net is trained to correctly predict the color and spatial map, while simultaneously modifying the SLM pixels to converge to the optimal pattern that balances the performance of color and localization determination (Supplemental Information).

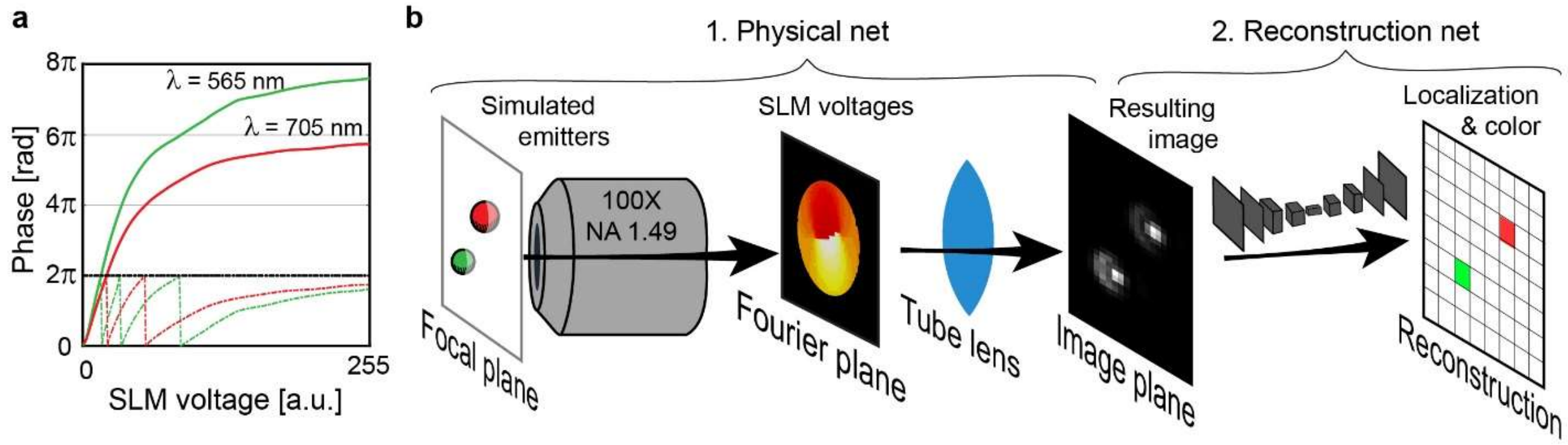

**Fig. 5** Design of an optimal SLM-pattern using neural networks. **a** An SLM imparts a chromatically-dependent phase delay as a function of applied voltage. **b** A schematic depicting the simulated process for generating an optimal phase mask consisting of 1. a Physical net used to generate the resulting PSFs for a particular SLM voltage pattern, and 2. a Reconstruction net, which decodes the generated images.

The net converged to a novel SLM-voltage pattern (Fig. 6a), which imparts two distinct phase-delay patterns depending on the wavelength (Fig. 6b), yielding two different PSFs, one for each color (Fig. 6c). To compare between the performance of the reconstruction net when using the optimal PSF versus the unmodified PSF, we repeated the Qdot experiment described earlier (Fig. 2). For each sample position, an additional image was taken with the optimal SLM pattern (when the SLM is inactive, no PSF-modification occurs). The experimental PSF resembles the simulations (Fig. 6d) and the differences in the intensity distributions are observable in the image cross-sections (Fig. 6e). The net correctly predicted the color of 99.4% ± 0.5% Qdots (N = 2195), improving over the already high 96.4% for the normal PSF (Fig. 6f).

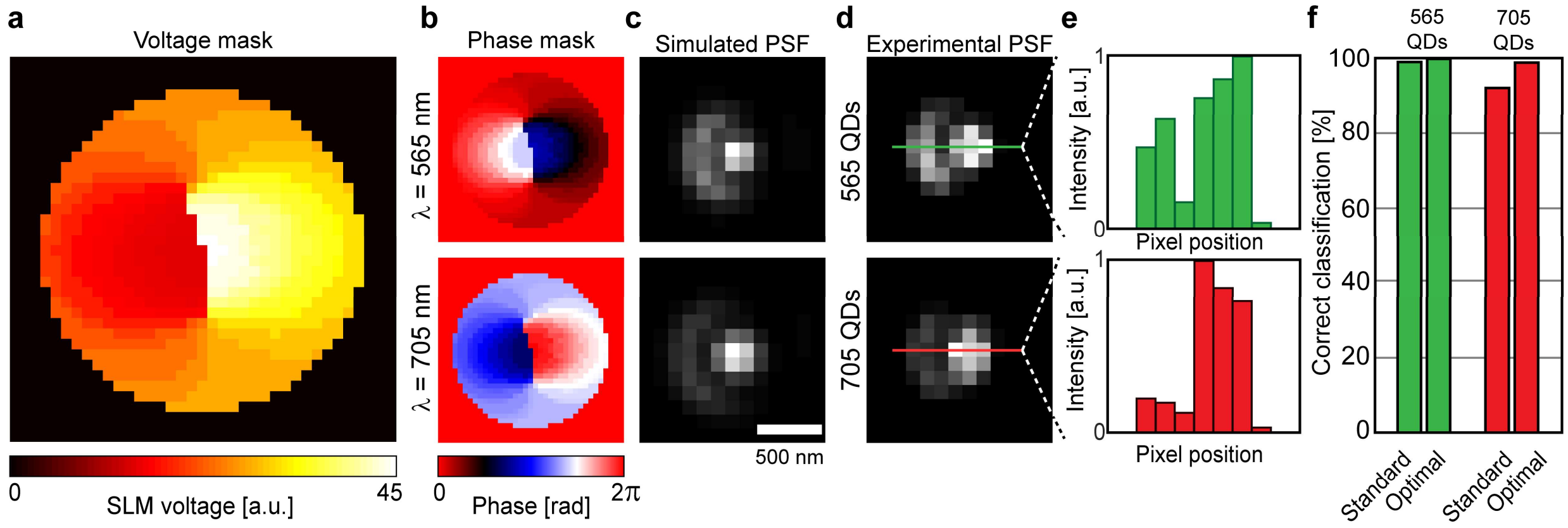

**Fig. 6** Optimal PSF engineering with an SLM**. a** The optimal SLM-voltage pattern for color determination by a neural net. **b** The phase delay imparted to 565 and 705 nm light. **c** Simulated PSFs. **d** The experimental PSFs measured with Qdots. **e** Pixel values of the cross section from experimentally measured PSFs. **f** Performance of color determination for the normal and optimized PSFs.

## Discussion

Neural networks have been shown to constitute a powerful tool in microscopy posed to replace existing algorithmic approaches[22,28,29]. Here, we have demonstrated how deep learning is capable of performing roles traditionally accomplished with physical components. Post-process, software-tools can be advantageous over hardware-based methods due to a lower implementation cost, system adaptability, and further optimization without the requirement of collecting new, experimental datasets.

Localization microscopy is particularly well suited for deep learning because emitters form a relatively homogeneous population. This reduces the requirement for enormous sets of diverse training data which is often required for deep learning. Historically, localization microscopy relies on fitting only a few parameters to an image of an emitter (e.g. amplitude, Gaussian widths, astigmatism angle, and background, or for maximum-likelihood-estimation – all of the pixels in the region of interest of the PSF[30]), whereas neural nets rely on orders-of-magnitude more tuned parameters. This flexibility of the network architecture allows it to capture more subtleties in the data. For example, in the Qdot experiment using the standard PSF, the net far outperformed the maximum-likelihood (matched filter) solution.

The strength of this process is apparent in the diffusing, fluorescent-microsphere experiment. Images of randomly moving particles can vary significantly due to the stochastic step size in diffusion. However, with relatively short trajectories, 100 frames or less, the net achieves a high classification rate. Interestingly, the net exhibited better performance on mobile beads than stationary ones, but still classified both robustly. This demonstrates the versatility of our approach and its applicability to tracking experiments in biological samples, where particles exhibit various, and sometimes switching, behaviors.

When applied to single-molecule blinking data, a number of new challenges are introduced. First, relative to Qdots and fluorescent beads, the SNR is low and heterogeneous. To achieve a sufficient SNR for color determination, we used molecules that had >2000 signal photons. While certain fluorophores, such as Alexa 647, exhibit spectacular blinking properties that make them well suited for this requirement, the choice of a second fluorophore is more difficult, and its behavior may be optimal under different blinking conditions[31]. To test the performance of the net at various SNR conditions, we simulated emitters on a noisy background and measured the color determination efficiency. We found that the net could identify emitter colors even in adverse conditions, but performed best at SNR > 4 (Fig. S5). In our experimental data, field-dependent aberrations may confound the net; however, larger training-set sizes may solve such issues. Current development of new fluorescent dyes and advances in imaging-buffer formulations will further improve the applicability of our method by improving the SNR and the spectral range of suitable labels.

Another area in which super-resolution algorithms are quickly improving is dealing with the *overlap* problem[22,32–36]. Briefly, single-molecule, super-resolution datasets consist of many frames containing unique sets of emitters. To reconstruct a super-resolved image, as many molecules as possible must be localized. Thus, for a given acquisition time, increasing the density of emitters in each image directly translates into a better reconstruction, so long as the emitters can still be localized with high precision. To test our method in varying-density environments, we simulated overlapping emitters and found that the classification net performed well up to ~6.2 emitters/$\mu m^2$, which is a medium-to-high density of emitters (Fig. S4).

As with any algorithm, some misclassification is inevitable, particularly at a low SNR. For the development of the optimal PSF, the net needs to balance two constraints: 1. the limited-signal requirement, where grouping the photons together improves the SNR, and 2. the color-determination requirement, where spreading the light differently makes the determination more efficient. The solution that the net found fulfills these two demands simultaneously. The difference between the maximal pixel value of a point source blurred by the optimal SLM pattern and the constant SLM pattern (diffraction limited blur only) is only 34% for the red PSF and 46% for the green PSF (Fig. 6). In other words, the optimal PSFs determined by the net are spread just enough for their shapes to be distinct from one another when observed through noise, so that color determination performance over the experimental data approaches 100%.

The resulting pattern achieves excellent performance even when the SNR is low. Since the PSF was devised with simulated data, the net stabilized on an SLM pattern in which the photon budget for the red and green emitters are similar, and thus no bias was introduced that gives preference to the red over the green emitters, or vice versa. At very-low SNR, the net is forced to prioritize grouping the photons, and only slightly changes the appearance of the emitters (Fig. S8).

Experimentally, we demonstrated color classification between two distinct emitters; however, in theory, the approach is not limited to any particular number of species, so long as their spectra are different enough. To evaluate the number of colors that could be distinguished using our method, we performed two types of simulations: first, we assessed the efficiency of the net to distinguish emitters with varying spectral differences (Fig. S6). While the net performs best when the spectra are well separated, we found that the net still performs well, greater than 90% success rate with no additional optics needed (the standard PSF), when the wavelengths were as close as 40 nanometers. When the optimal phase mask is introduced, even extremely small differences can be measured given a sufficiently high signal-to-noise ratio, such as those observed in the Qdot experiment. Experimentally, such small differences are typically impractical due to spectral broadening; however, in single-molecule experiments at low temperature, such a regime has been observed[37–39].

Evaluating the difference between two spectrally-similar emitters is still a different problem than categorizing more colors. Using our optimal PSF-engineering algorithm, we compared the performance of the method on simulated image data containing 2, 4 and 5 different colored emitters, matching those available for commercial Qdots (Fig. S7). Using single-images containing a mix of PSFs, the net performance barely degrades with 4 different emitter types (>99% correct color determination), and still achieves greater than 91% classification with all 5.

Here we have shown an implementation of a useful architecture for discrimination of different PSFs, namely in color. The technique, however, could also be used for analyzing any other effecters on the shape of the PSF (e.g. z-position, molecular orientation, movement dynamics, number of contributing emitters, etc.). To optimally discriminate between PSFs, we have shown that PSF-engineering can be done in coordination with net training to maximize on the strengths of the reconstruction net, which do not follow the same process as most-likelihood estimators.

## Methods

### *Deep learning*

Localization of emitters was performed either using the ImageJ[40] plugin ThunderSTORM[41] or a custom peak-finding implemented in algorithm in MATLAB (Mathworks).

The deep neural nets were implemented in MATLAB and several functions of the MatConvNet package[42]. Our setting required two types of nets: one for color determination, and one for simultaneous optimal PSF design and color determination with the resulting optimal PSF. Color determination is in fact a conventional classification task. For our color determination net, we adopted an architecture similar to that proposed by Ledig et al.[43] Our net contains 9 convolutional layers with an increasing number of channels and a decreasing spatial resolution (implemented by three stride 2 convolutions), followed by two dense (fully-connected) layers and a sigmoid layer that outputs a color probability. We used the cross-entropy loss and Adam optimization for training this net[44].

Determining an optimal microscopic PSF through the design of an SLM phase pattern, is a unique task that has not been addressed in previous work. Here, our architecture comprised a physical (sub-) net whose role is to simulate the image generation process with an SLM, and a reconstruction (sub-) net whose role is to classify colors based on the PSFs induced by the learned SLM. Both nets were trained simultaneously so as to learn the optimal voltages of each of the SLM's pixels and to subsequently perform color discrimination. In neural net terminology, the SLM's voltage mask can be thought of as a nonlinear layer, whose parameters have to be optimized, similarly to conventional linear layers.

The physical sub-net selects the voltage value for each pixel among 50 possible voltage values, similarly to the principle used by Chakrabarti[19]. The induced red and green PSFs are then convolved with simulated red and green point sources, respectively, and the resulting grayscale image is subsampled and contaminated by Poisson noise.

The color reconstruction sub-net contains eight convolutional layers, followed by two deconvolution layers, each doubling the spatial resolution (See Supplementary Information). It also contains a skip-connection branch with one deconvolution layer that increases the spatial resolution by 4 and bypasses nine of the convolutional layers of the main branch. This is done in order to alleviate the gradient-vanishing problem[45]. The outputs of both branches enter a last convolutional layer followed by a sigmoid layer. The net's output is a probability map for each pixel of the gray image being red or green on a high resolution grid. The loss is again cross-entropy. Additional details on the net architecture and training are in the Supplementary Information.

*Microscopy*

Imaging experiments were performed using a standard inverted microscope (TI Eclipse, Nikon), equipped with an XY Proscan III translational stage and a Nano-Z Piezo stage (both Prior Scientific). The instrument was controlled with Nikon Imaging Software and illuminated with a fiber-coupled laser-light source (iChrome MLE, Toptica). To allow for the placement of additional optics (*i.e.* the spatial light modulator), the imaging path was extended with two 15 cm lenses (Thorlabs). All images were recorded on a high quantum-efficiency sCMOS camera (95B Prime, Photometrics). Microscope configuration schematics are shown in Fig. S10.

For Qdot experiments, 565 and 705 nm emission-peak nanoparticles (Life Technology) were diluted in 1% PVA and spin coated onto a No. 1.5 coverslip slide (Thermofisher) achieving a final density of 0.08/$\mu m^2$ green Qdots, 0.05 red Qdots/$\mu m^2$. Samples were then excited with 405 nm light and imaged through a Nikon 100X NA 1.49 TIRF objective in Epi-illumination mode. The emission light was chromatically filtered with a dichroic and long pass filter (ZT488rdc & ET500LP, Chroma) to remove background and scattered illumination light. To distinguish the true color of the Qdots recorded with the grayscale camera, an additional 565/70 or 650 LP filter (both Semrock) was inserted in the imaging path. In Qdot experiments with an optimized point-spread functions for color determination, a liquid crystal spatial light modulator (SLM, PLUTO-VIS, Holoeye) was used at a position conjugate to the back focal plane of the objective, after a linear polarizer. For the matched-filter comparison, the mean PSF was generated by subpixel-shifting each ROI containing a Qdot according to the localized position in a field of view, normalizing each image and taking the average. In the dataset, each PSF was then shifted according to its localized position and the correlation with the mean red and green PSFs were compared.

For the diffusing-bead experiment, 100 nm green and 200 nm red fluorescently labeled microspheres (Life Technology) were diluted into 40% glycerol in water (v/v). From the mixture 10 uL was then pipetted onto a glass coverslip and pressed onto a glass slide and sealed with clear nail polish. Both surfaces were pretreated with a ~20 mg/mL casein solution to decrease the propensity for sticking of the fluorescent beads to the glass. In most regions of the sample, the beads remained in solution for several hours; however, there were areas of the sample where the passivation layer was flawed and the majority of beads had adhered to the surface. Three fluorescent filters were used in combination with different excitation-laser combinations to image the green, red, and both beads at once. Green-bead images were recorded with a 488 nm excitation and a green filter set (ZT488rdc & ET500lp, EM525/50bp, Chroma); red beads were imaged with a 650 nm laser and a red filter set (ZT650rdc, EM650lp); images of both beads were done using a multi-bandpass filter set (ZT405/588/561/647rpc, ZET405/488/561/647m, Chroma). All imaging was done using a Nikon 20X air objective, NA 0.95 without the additional 4f extension used in the Qdot experiments.

Cell experiments were performed on fixed and antibody-labeled HeLa cells. Cells were cultured on glass-bottom culture dishes for 48 hours before fixation and labeling. Fixation was performed following a modified paraformaldehyde protocol outlined by Whelan et al[46]. Briefly, cells were treated with ice cold 3.7% paraformaldehyde for ten minutes, permeabilized with 0.25% Triton X in 0.01M (1X) PBS, quenched with 0.3M glycine in PBS, blocked with 10% goat serum, 3% BSA, 0.3M glycine, and labeled in blocking media containing Alexa-647 conjugated anti-tubulin and Alexa-555 conjugated anti-TOM20 antibodies (both Abcam) for 8 hours at 4 degrees. Cells were then washed with PBS 5X with increasing incubation periods from 30 seconds to 10 minutes. After washing, an additional fixation step with 3.7% paraformaldehyde was added, followed by a 10 minute treatment with 0.3M glycine in PBS and 3X additional washing steps with PBS. For imaging, the media was replaced with a glycerol-based imaging buffer suitable for both Alexa647 and Alexa-555[47]: 95% glycerol (w/w), 10% glucose (w/v), 50 mM TRIS supplemented with a commercially available cysteamine-based blinking supplement containing oxygen scavengers (Abbelight). Imaging was performed with total-internal-reflection (TIR) illumination using 405 nm reactivating light, 488 and 561 nm light to pump Alexa 555 and 647 nm light to pump Alexa 647. Analysis was performed using custom MATLAB localization routines that combined local ROIs from sequential frames in which the same molecule was active.

Code and data will be made available.

**Supplementary Information**

# Table of Contents

---

## *1. Architecture and hyper parameters of the color determination net*

The architecture and training procedures for the color-determination net were similar for all experiments using the standard PSF (Fig. S1).

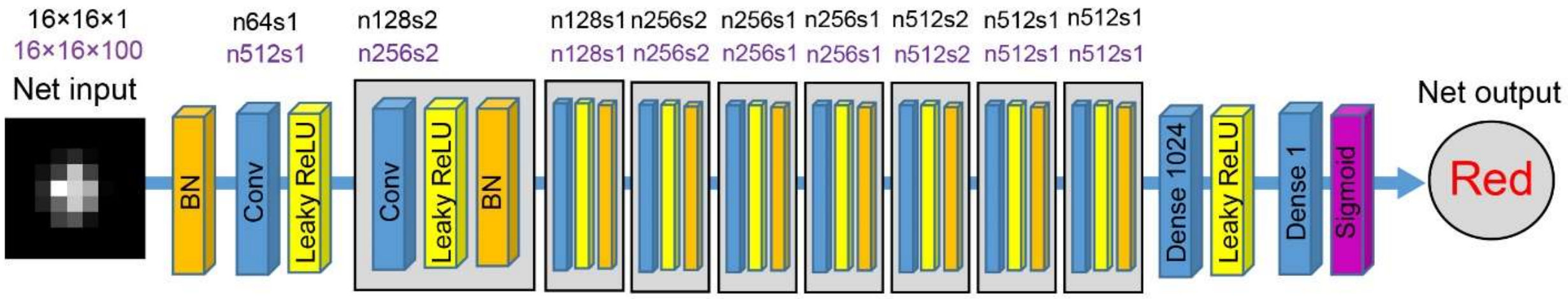

**Fig. S1** Architecture of the color determination net. Top: the number of feature maps (n) and stride (s) of each convolutional layer. Black denotes the Qdot net for the standard PSF and the super-resolution net. Purple text describes the net for moving beads.

The first step in training the net is to obtain a suitable image library. Patches containing a small number of pixels around an emitter are extracted from the image data $(11 \times 11)$. A randomly selected fraction of these patches are used in each training iteration as is, while the remaining patches go through an augmentation process to ensure that the net remains broadly applicable to varying SNR conditions.

For the motionless emitter nets with the standard PSF and the optimal SLM patterns, the training data was split into two even parts. For the augmented half, the median background gray level of each patch was subtracted and a random background floor level in the range [400, 800 DU] was added together with a Poisson noise with a random variance, λ, in the range of [16, 64].

For the moving emitter net, 100 out of 500 patches were randomly selected for background augmentation. The median background gray level of each patch was subtracted and a random background floor level in the range [150, 350 DU] was added together with a Poisson noise with a random λ in the range of [25, 121].

For the single-molecule net, patches of blinking emitters with $SNR > 8$ were selected from the training data. For the augmented half, the median background gray level of each patch was subtracted and Poisson noise with a random variance, λ, in the range of [0, 250] was added.

For all three nets, the patches were randomly shifted in the horizontal and vertical directions in the range of [-3, 3] pixels and a $16 \times 16$ patch was obtained by replicating the borders of the $11 \times 11$ patch.

The detailed architecture of the color determination net is shown in Fig. S1. The first Batch-normalization layer in the net is used to normalize the data. Its gains and biases are set to ones and zeros and are not learned parameters.

All the convolutional weights of the net are $3 \times 3$ in size.

A dropout with $p = 0.5$ is implemented after each Leaky ReLU layer ($\alpha = 0.01$), except for the last one, and L2 regularization is used with $\lambda = 10^{-6}$.

The Adam optimizer is used with $\beta_1 = 0.9$, $\beta_2 = 0.999$, and $\varepsilon = 10^{-8}$ to update the net's parameters.

We use a cross-entropy Loss:

$$LOSS = -\frac{1}{10 \cdot N} \sum_{n} [GT \cdot log(Z + \varepsilon) + (1 - GT) \cdot log(1 - Z + \varepsilon)] \quad (1)$$

Where $n$ is the batch image index, $N$ is the batch size, $GT$ is the ground truth ($0/1$ for a red/green emitter) and $Z$ is the net's output.

The Qdot net using the standard PSF was trained for 28K iterations with a learning rate of 0.0001 and a batch size of 32, and for additional 476K iterations with the same batch size and a learning rate of 0.00001.

The Qdot net with the optimal SLM pattern was trained for 98K iterations with a learning rate of 0.0001 and a batch size of 32, and for additional 30K iterations with a batch size of 64 and the same learning rate.

The diffusing-bead net was trained for 105K iterations with a learning rate of 0.0001, for 20K iterations with a learning rate of 0.00001 and for 50K iterations with a learning rate of 0.00005, all with batch size 32.

The single-molecule net was trained for 158K iterations with a learning rate of 0.0001 and batch size 16.

2. *Architecture and hyper parameters of the optimal SLM estimation net*

*The Physical Net*

The physical net architecture is presented in Fig. S2. The net's purpose is to simulate imaging emitters at randomly placed positions after encountering an SLM with a particular phase pattern. The SLM voltage weights are the only learned parameters. There are $215 \times 215 \times 50$ parameters, where $215 \times 215$ represents simulated area of the back focal plane, and the depth represents all the possible SLM voltages. Note that we only use a small range of voltages in the range [12, 61V] out of the full possible SLM voltage range of [0, 255 V], for improved correspondence between simulation and experiment. These weights are then multiplied by a scalar parameter, $\alpha$, which is slowly increased as iteration number increases, according to:

$$\alpha(t) = 1 + \gamma * t^2 \qquad (2)$$

where $t$ is the current iteration number and $\gamma = 2.5 \cdot 10^{-5}$.

The result is passed through a Softmax layer, which normalizes it such that each one of the $215 \times 215$ SLM pixels contains a probability vector for each one of the possible 50 SLM voltages. The role of $\alpha(t)$ is therefore to make the probabilities sharper as the iteration number increases. Then, an inner product is implemented between each one of the voltage probability vectors and a vector of the corresponding voltages. This operation chooses one voltage value for each SLM pixel, and the result is a $215 \times 215$ SLM voltage pattern. The SLM pattern is then split between the red and green PSFs channels.

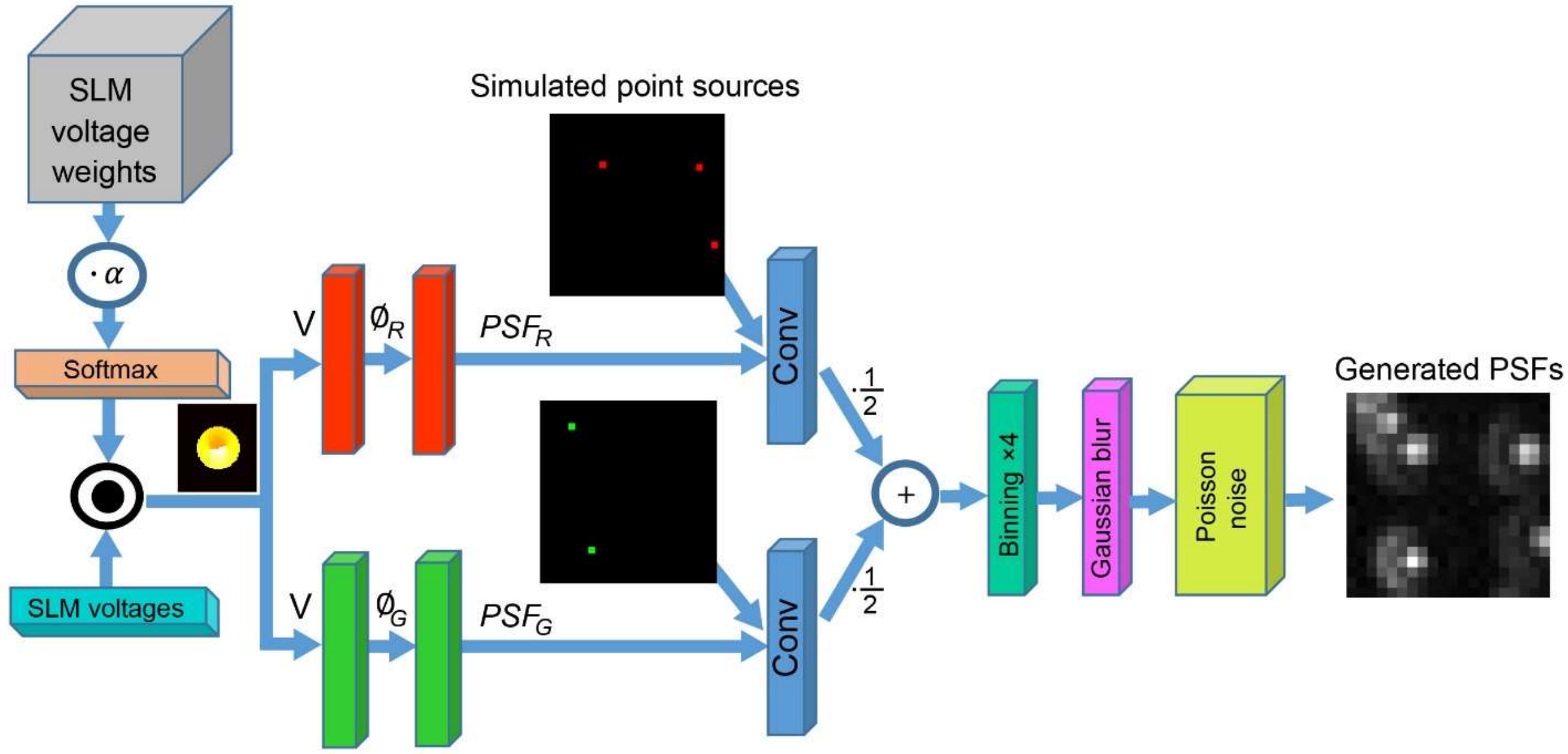

**Fig. S2** The Physical network architecture for the optimal SLM estimation.

*The green PSF channel:*

The imparted phase for green radiation ($\lambda = 565\ nm$), is obtained from the voltage pattern by using the SLM response curve (Fig. 5a). The green PSF is then produced as

$$PSF_G = \left|\mathcal{F}^{-1}\{circle_G \cdot e^{i\cdot\emptyset_G}\}\right|^2 \tag{3}$$

where $PSF_G$ is the green PSF, $\mathcal{F}^{-1}$ represents the 2D spatial inverse Fourier transform and $circle_G$ is a centered circle. This circle is present due to the 'cone' of light rays that are collected by the microscope from an emitter and projected onto its pupil plane and then on the Fourier space of the 4f system. The physical size of the circle in the SLM Fourier plane is given by

$$D = 2f \cdot \frac{NA}{M} \tag{4}$$

where $D$ is the diameter of the back focal plane image, $f$ is the focal length of the first lens in the 4f system, $NA$ is the numerical aperture of the objective, and $M$ is the image magnification. For simulations we use a high resolution grid with a $4X$ reduced pixel size relative to the camera's pixel size. The Fourier space size in simulation can be obtained by:

$$A_G = \frac{\lambda_G \cdot f}{pixel_{HR}} \tag{5}$$

Using all the physical sizes of our optical system, $f = 150mm$, $NA = 1.49$, $M = 100$, $\lambda_G = 565nm$ (mean wavelength for the green Qdots used in experiments), $pixel_{HR} = 2.75\mu m$, we can confirm that the circle diameter is 14.5% out of the green SLM space size $\left(\frac{D}{A_G}\right)$, or 31 pixels out of the 215 green SLM size that was chosen arbitrarily in the simulation.

*The red PSF channel:*

The red PSF is produced similarly to the green as

$$PSF_R = \left|\mathcal{F}^{-1}\left\{circle_R \cdot e^{i \cdot Padding(\emptyset_R)}\right\}\right|^2 \tag{6}$$

Using $\lambda_R = 705nm$ (mean wavelength for the red Qdots used in experiments), we can confirm that the circle diameter (which is independent on the wavelength) is 11.6% out of the red SLM space size $\left(\frac{D}{A_R}\right)$, which means that the red SLM space size in the simulation is $\frac{31}{0.116} = 267$ pixels. The padding in (Eq. 6) is a zero padding operator that resizes the $215 \times 215$ $\emptyset_R$ phases to $267 \times 267$.

Next, the net produces a high resolution gray image as

$$Gray_{HR} = \frac{PSF_R * Sources_R + PSF_G * Sources_G}{2} \tag{7}$$

where $Sources_{R\,\&\,G}$ are the high resolution grid images of the red & green emitters' locations and '$*$' denotes convolution.

We use in the simulation $30 \times 30$ detector's grid patches in which a random number between [5, 10] red and green points are located in random positions over the $120 \times 120$ high resolution grid. Each point is assigned with a random signal value between [6500, 13000 DU].

$Gray_{LR}$ is obtained by passing $Gray_{HR}$ through a binning operation (performed by a $4 \times 4$ mean pooling layer multiplied by 16) and a convolution with a Gaussian with a standard deviation of 0.5 pixel that simulates a mild optical blur in the optical system.

The last stage is obtaining the images by adding a background and contaminating with Poisson noise:

$$Image = Poissrnd\{\lambda_{Poisson}\} = Poissrnd\{Gray_{LR} + background\} \tag{8}$$

where the background term is a constant image with a random gray level in the range [144, 676]. The typical SNR in simulated emitters is around 30, but goes as low as 13.

An emitter's SNR is defined as:

$$SNR = \frac{\Delta Signal}{std(Noise)} = \frac{Signal_{max} - background}{\sqrt{background}} \tag{9}$$

Eqn. 8 is the input to the reconstruction net.

*The Reconstruction Net*

The reconstruction net architecture is presented in Fig. S3b. The input to this net is the low resolution $30 \times 30$ image of generated PSFs. Its purpose is to generate the high resolution $120 \times 120$ localizations and color determinations maps.

The batch size is 16 and the Adam optimization is used with the same parameters as in the color determination net. No regularization is used.

We use a weighted cross- entropy loss

$$LOSS = -\frac{1}{10 \cdot N} \sum_{h,w,d,n} [Mask \cdot GT \cdot log(Z + \varepsilon) + Mask \cdot (1 - GT) \cdot log(1 - Z + \varepsilon)] \tag{10}$$

where $n$ is the batch image index, $N$ is the batch size, and $GT$ is the ground truth: a $120 \times 120 \times 2$ image for one batch image. The first layer is a $120 \times 120$ grid of red predictions, consisting of 1 or 0 values that denote the existence of an emitter in each pixel, and the second image is the green predictions. $Z$ is the net's output, and the sum is over all of its dimensions.

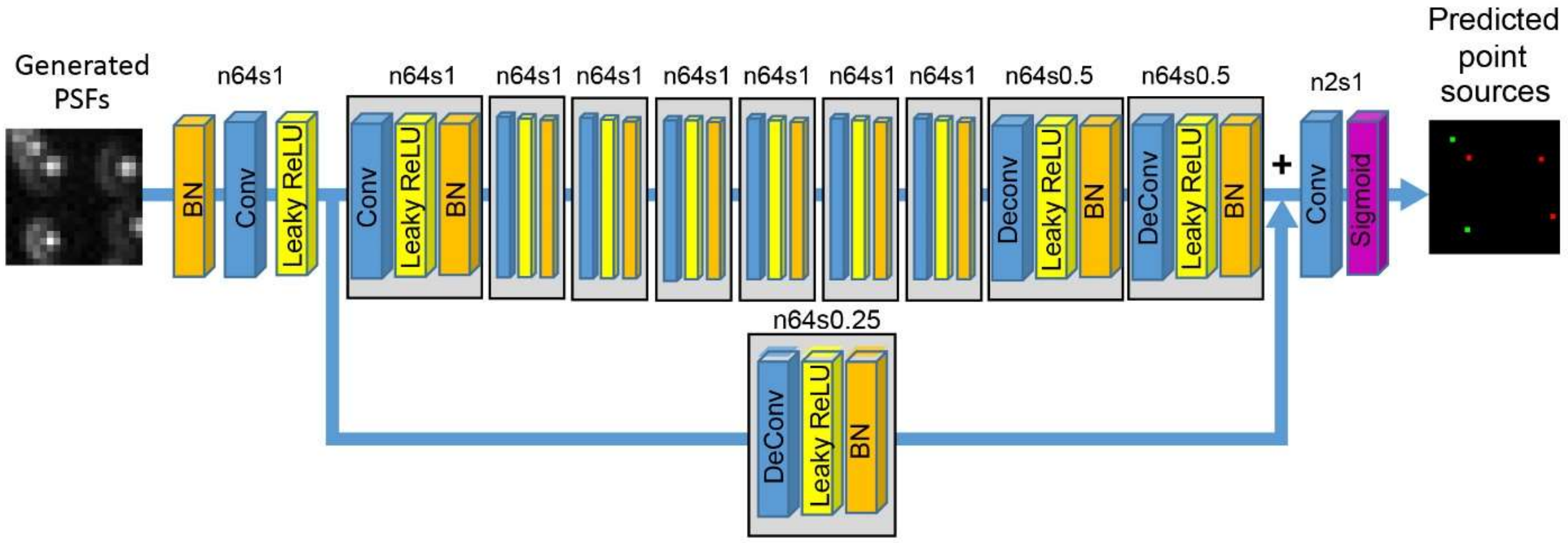

**Fig. S3** The Reconstruction network architecture for the optimal SLM estimation.

The mask is a $120 \times 120 \times 2$ image of scores. Its purpose is to encourage the net to correctly predict pixels that contain emitters by assigning them a higher score. Producing the $Mask$ involves assigning each emitter in the red emitters' locations image a score of $\frac{3}{N_{red}}$, where $N_{red}$ is the total number of red emitters, and each emitter in the green emitters' locations image a score of $\frac{3}{N_{green}}$, where $N_{green}$ is the total number of green emitters. Then, the two images are arranged to produce Mask. Finally, all the pixels in Mask that are zeros are set to $\frac{1}{N_{zeros}}$, where $N_{zeros}$ is the total number of zeros in Mask. This process provides the net a total score that is 3 times greater when it correctly predict all the red and all the green emitters instead of correctly predicting all the pixels that don't contain emitters. Finally, the pixels in Mask that are outside the central $80 \times 80$ central pixels are set to zero, so emitters near the borders do not contribute to the loss.

The optimal SLM estimation net was trained for 270K iterations with a learning rate of 0.0001.

### *3. Evaluation of the net under various conditions*

To assess the net's performance in a variety of experimental-like conditions, images were simulated with various emitter densities (Fig. S4) and SNRs (Fig. S5). For these simulations, we fix the optimal SLM pattern in the optimal SLM estimation net and update only the reconstruction net's weights. The reconstruction net is trained with a random number of red emitters between [2, 20] and a random number of green emitters in the same range. We also assign random signal values for red and green emitters in the range [12000, 24000 DU]. All other net's parameters are the same as in section 2 We initialize the reconstruction net weights to the weights learnt in section 2 and train for 100K iterations with a training rate of 0.00005.

*Testing emitter densities*

For each density of emitters, 100 simulated images are generated with a random number of red and green emitters (the total number of emitters corresponds with that density). Next, the reconstruction net is used in order to generate localization and color determination maps. Finally, we perform a post processing which leaves in those maps only the pixels that are local maxima, threshold the result at 0.95, calculate the Detection/Color Determination and False Alarms measures and average them over the 100 examples.

The *Detection* metric is the percent of color detected emitters out of all emitters (Fig. S4a), even if their color determination was wrong.

The *Color-Determination* metric is the percentage of detected emitters, whose color was determined correctly.

The *False Alarm* measures the percentage of falsely predicted emitters where none were present.

An example region with a 6.2 $\left[\frac{emitters}{\mu m^2}\right]$ density (30 emitters in a $2.2 \times 2.2\ \mu m^2$ sample region, which correspond to a $20 \times 20$ pixel$^2$ ROI area on the detector) is shown in Fig. S4d-f.

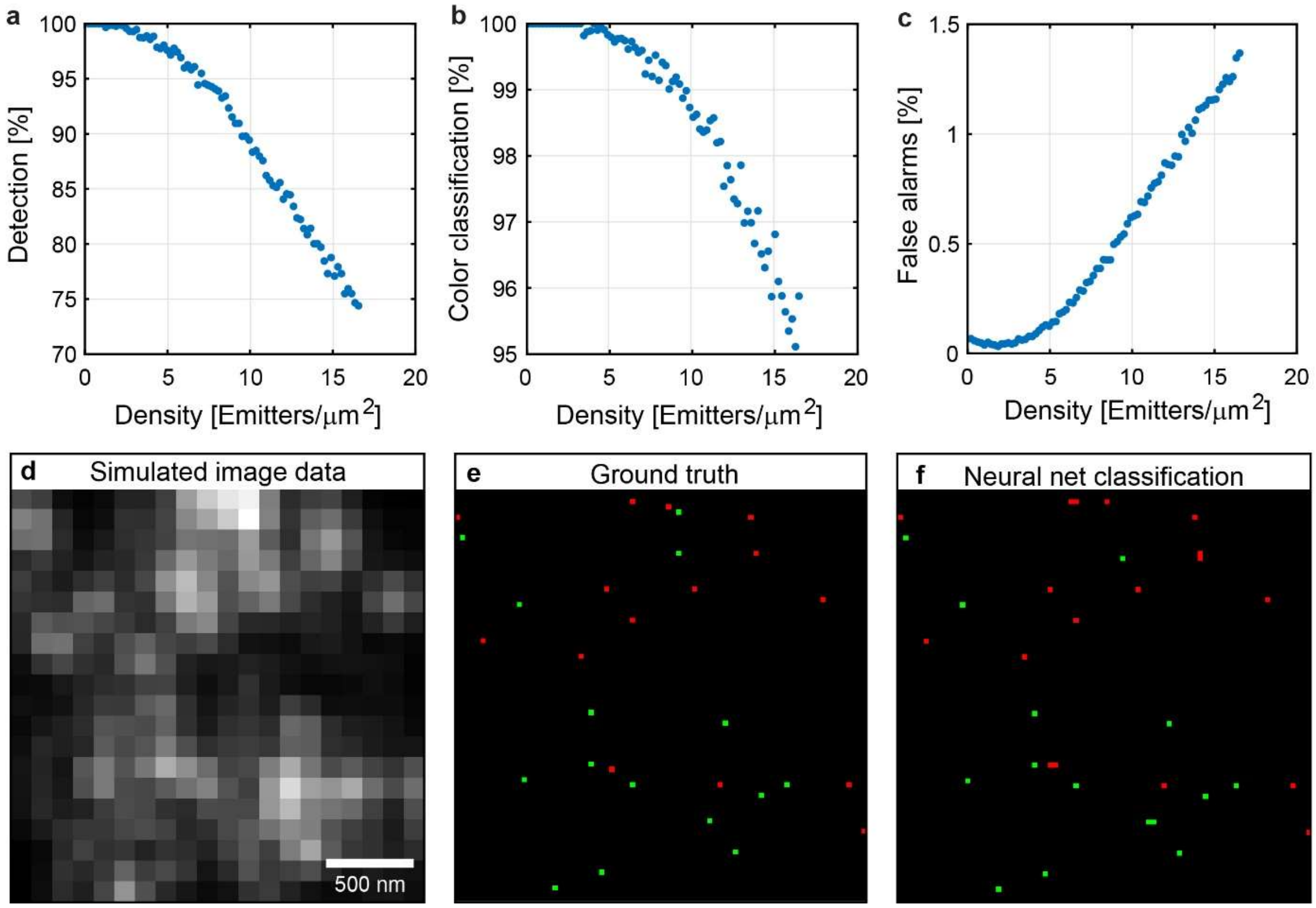

**Fig. S4** Evaluating performance over different emitter densities. **a** Detection performance. **b** Color determination performance. **c** False alarms. **d** An example of a generated PSFs image with a density of 6.2 $\left[\frac{\text{emitters}}{\mu\text{m}^2}\right]$. **e** The ground truth of d. **f** The net's prediction of d.

*Testing emitter SNRs*

For each SNR, 100 images are generated with a Poisson noise, $\lambda = 410$, and a random signal value in the range [1400, 8400 DU]. One red and one green emitter is placed in each image. The reconstruction net is used in order to determine the localization and color of identified emitters in the FOV.

The measures and an example for an SNR of 4.3 are shown in Fig. S5.

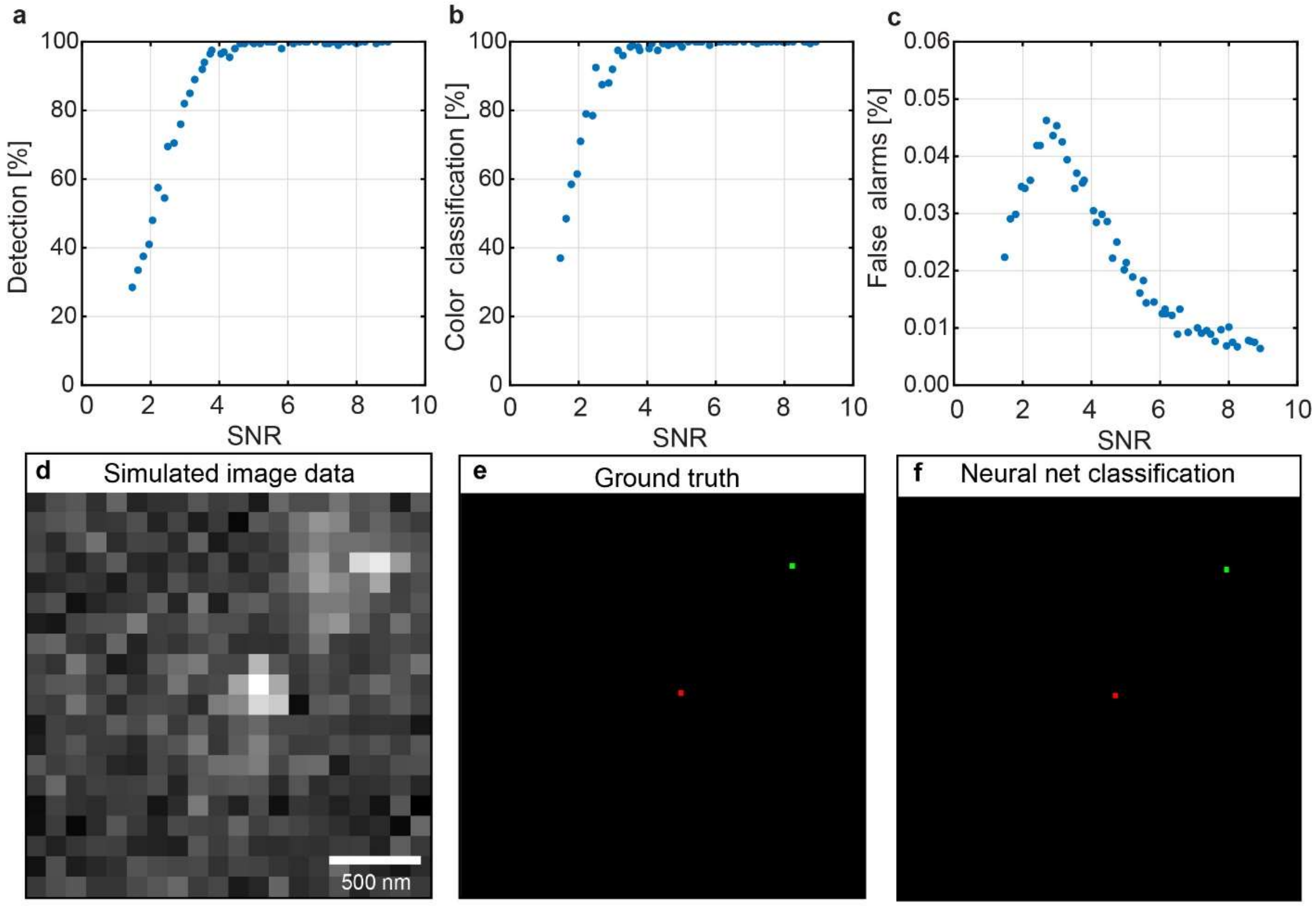

**Fig. S5** Evaluating performance over different emitters SNRs. **a** Detection performance. **b** Color determination performance. **c** False alarms. **d** An example of a generated PSFs image with an SNR of 4.3. **e** The ground truth of d. **f** The net's prediction of d.

Using simulations, we determined the net's ability to distinguish emitters of various wavelength differences (Fig. S6) and more than two species (Fig. S7). First, we trained three different optimal SLM estimation nets with similar parameters to those we used so far, for three wavelength proximities:

$$\Delta\lambda_A = 5nm\ (\lambda_1 = 565nm,\ \lambda_2 = 570nm)$$

$$\Delta\lambda_B = 40nm\ (\lambda_1 = 565nm,\ \lambda_2 = 605nm)$$

$$\Delta\lambda_C = 140nm\ (\lambda_1 = 565nm,\ \lambda_2 = 705nm)$$

We also trained the reconstruction net with a constant SLM (no pattern) for emitters with the same proximities. For each simulation, we evaluated the average over 16K randomly generated images using the protocol described previously.

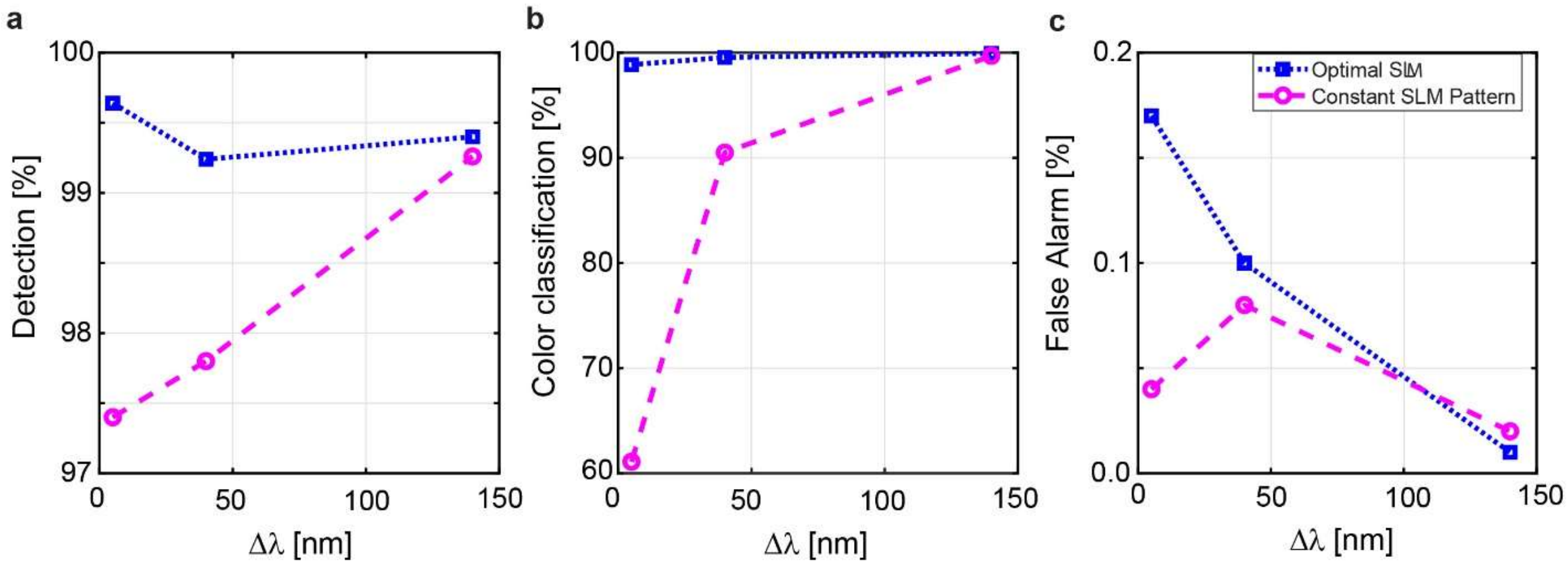

**Fig. S6** Evaluating performance over emitters' wavelengths proximity. **a** Detection performance. **b** Color classification performance. **c** False alarms.

*Evaluating net's performance over different number of emitters' colors*

To determine the number of spectrally distinct emitter types, $N$, that could be simultaneously identified using an optimized SLM pattern, we applied the following modifications to the optimal SLM estimation net described earlier: The SLM voltage is split into $N$ color channels, each one of them generates one color PSF. The image is then obtained by:

$$gray_{HR} = \frac{1}{N} \cdot \sum_{n=1}^{N} PSF_i * Sources_i \quad (11)$$

In the reconstruction net, the weight of the last convolutional layer consists of N filters, which means that the net's output has the size of $120 \times 120 \times N$, where $GT$ and $Mask$ are of the same size.

We used $N = 2, 4, 5$. One or two emitters per color are simulated per image. The 2 color net was described previously. The 4 and 5 channels nets are assigned a random signal value between [30K, 60K DU] and the colors are [545nm, 585nm, 625nm, 705nm] for the 4 channels net and [545nm, 585nm, 625nm, 655nm, 705nm] for the 5 channels net, corresponding to commercially available Qdots (Invitrogen). All the other parameters are the same as described previously. The 4-channel net was trained for 50K iterations with a learning rate of 0.0001 and a batch size of 8, and for 40K more iterations with a learning rate of 0.00005 and a batch size of 16. The 5-channel net was trained for 65K iterations with a learning rate of 0.0001 and a batch size of 8, and for 25K more iterations with a learning rate of 0.00005 and a batch size of 16. The performance was evaluated with 16K randomly generated images. In the 2 channel case, emitters were assigned a random signal between [18K, 36K DU], and in the 4 & 5 channel cases, emitters are assigned a random signal between [30K, 60K DU].

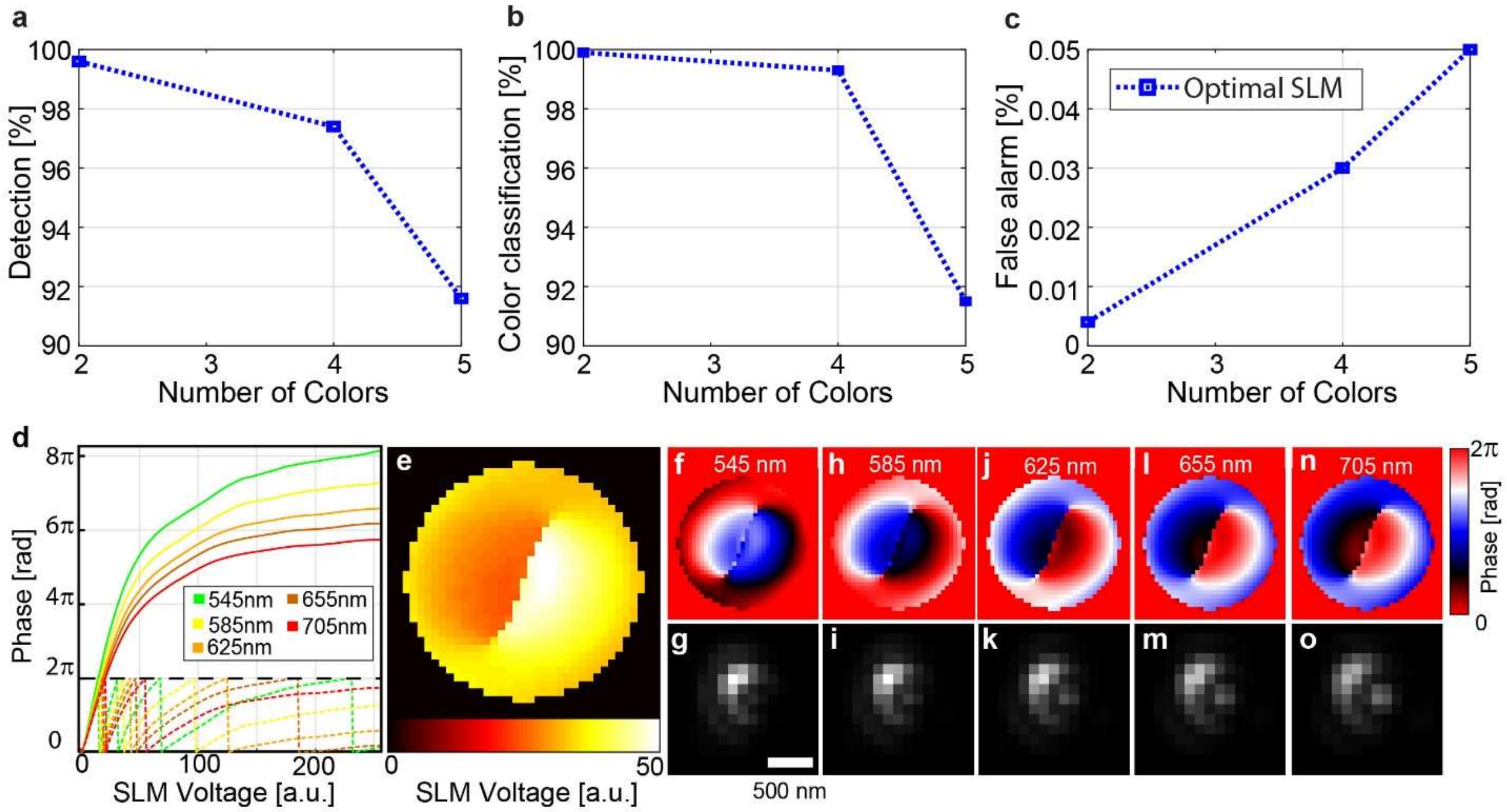

**Fig. S7** Evaluating performance over different numbers of emitter colors. **a** Detection performance. **b** Color determination performance. **c** False alarms. **d** The LC- SLM pixel's phase response as a function of input voltage for 545, 585, 625, 655, 705 nm. **e** The optimal voltage pattern. **f-o** The phase and simulated PSFs of each one of the five colors.

In cases when the SNR is very low, the optimal SLM only changes the PSF slightly from the normal one (Fig. S8), exhibiting mostly a linear phase ramp resulting in a non-informative lateral shift, along with a slight non-linear phase pattern. To find voltage pattern, we assigned each emitter fed to the physical net with a random signal value between [2400, 4800 DU] (40% of the signal values we used in Fig. 6).

After training the net for 150K iterations, the received measures are: detection 95.5%, color classification 98.8% and false alarm 0.1%.

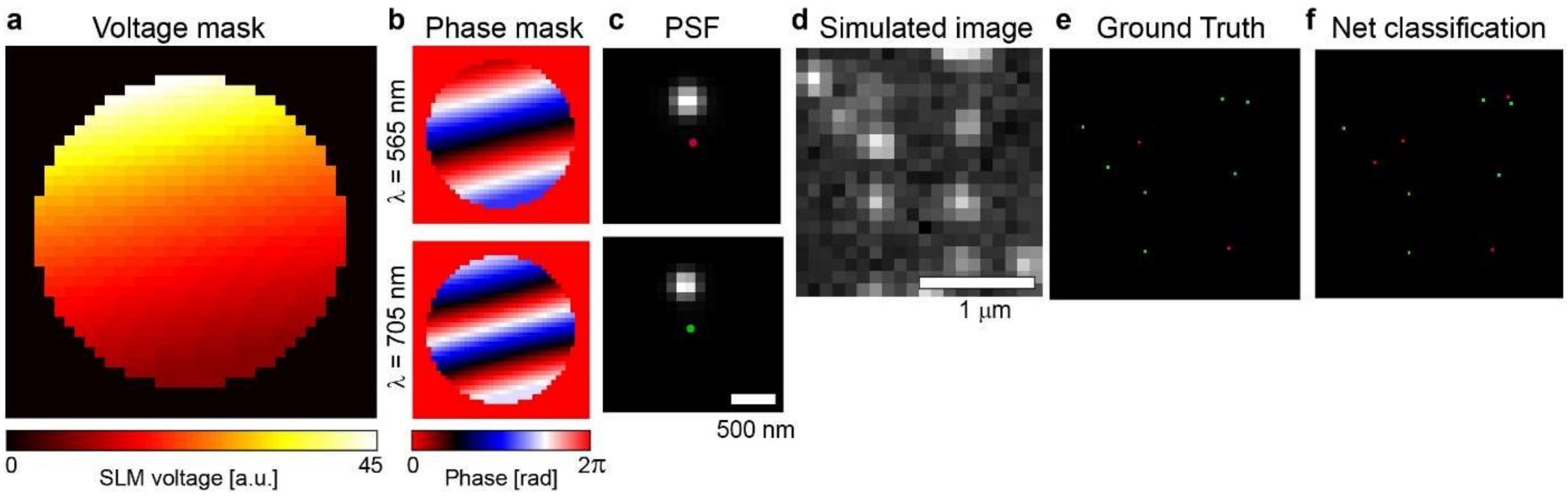

**Fig. S8** Optimal SLM results for low SNR conditions. **a** The optimal voltage pattern. **b** The red and green phase patterns. **c** The simulated red and green PSFs. **d** An example of a generated PSFs image with a low SNR. **e** The ground truth of d. **f** Classification by the net.

### 4. *Single-molecule localization microscopy*

The performance of the color-identification net depends on the factors described in previous sections. For single-molecule localization microscopy (SMLM), the limited number of useful emitters and limited SNR poses a challenge. To quantitatively assess the applicability of our method to multicolor SMLM, we generated movies of simulated blinking emitters using the Test-STORM toolbox[1] with a known ground truth. In simulations, two fluorophores were simulated: Alexa 647 and Alexa 568 (Fig S9a,b). All parameters used for simulations were the defaults of the program, except that for our analysis, two separate movies were combined so that the background was doubled so that the mean background and variance was $400$. The reconstruction net was trained as described previously for the super-resolution net used to analyze Fig. 4, except that 9915 red and $45215$ green emitters were used for training over $80K$ iterations with a learning rate of 0.0001 and batch size 16.

For quantifying the performance, an image was generated to contain spatially-separated fluorophores (Fig. S9c,d). The reconstruction net scores each localization $[0,1]$, which is then used to classify the localizations (Fig. S9e). The threshold to classify red and green emitters is a tunable parameter that should be weighted such that the misclassification rate is roughly even. Using a threshold of $0.95$ ($0.35$ was used for Fig. 4), the spatially separated localizations were used to quantify the correct classification rate. For red emitters it was $91\%$ and for green it is $86\%$ (Fig. S9f). This threshold was then used for a more qualitative evaluation on the remaining test datasets. Various overlapping structures were simulated (Fig. S9g). In the histogram, each pixel was colored given a red and green intensity based on the number of localizations of each red/green classified emitter (Fig. S9h).

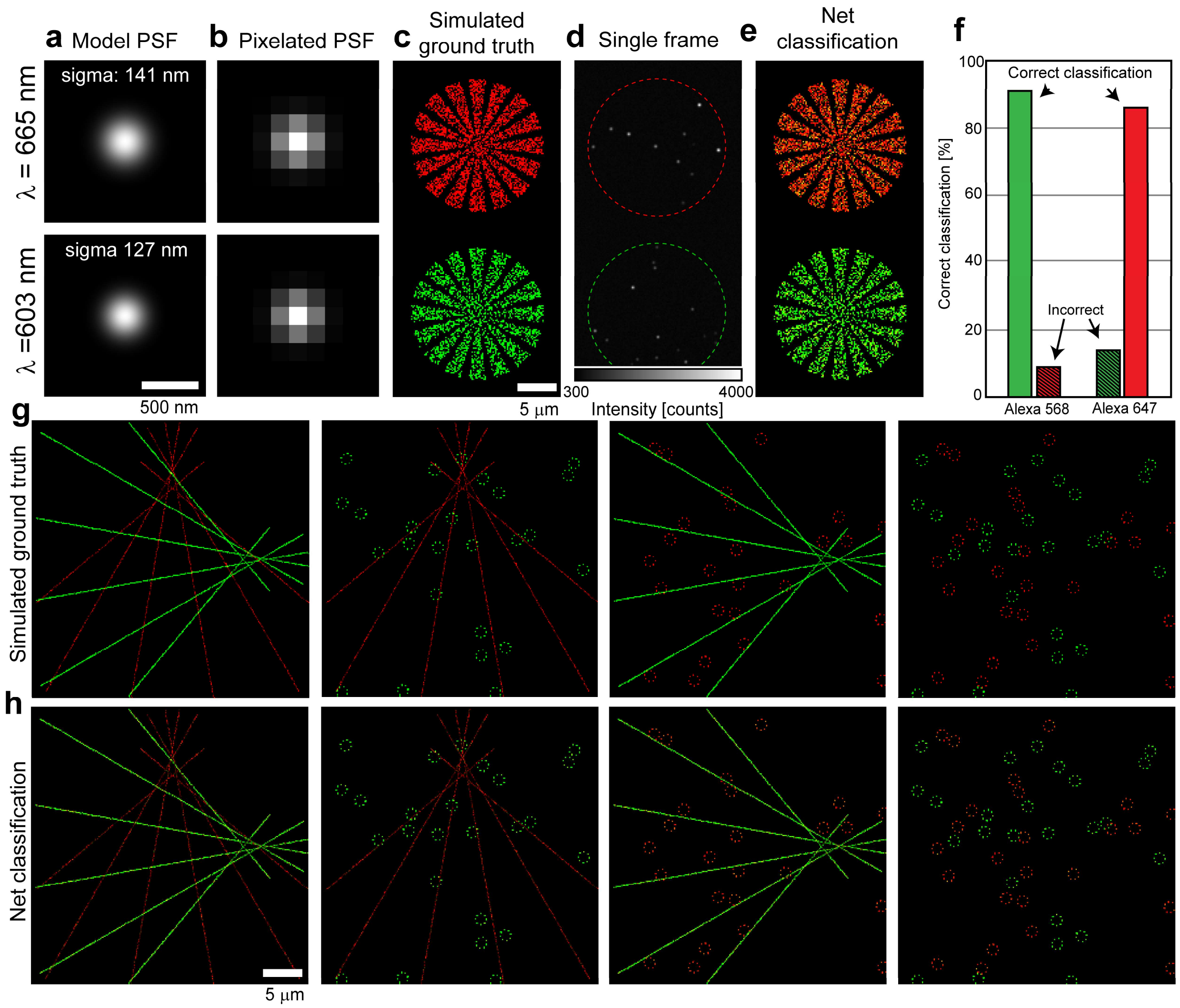

**Fig. S9** Simulated single-molecule data using testSTORM. **a** Simulated PSFs for Alexa 647 and Alexa 565. **b** Pixelated PSFs. **c** Reconstruction of two species-specific objects containing 25K fluorophores, each. **d** Representative single frame. **e** Pixels colored by the net according to density of emitters. **f** The fractions of correctly classified emitters by the net. **g** Ground truth and **h** net-classified super-resolution reconstructions.

### 5. *Microscope setup*

Two microscope configurations were used for experiments and simulations (Fig. S10). For the diffusing microsphere experiment and all simulations with the standard PSF the standard microscope configuration was used/simulated (Fig. S10a). For all other experiments and simulations, the 4f-extended configuration was used (Fig. S10b), where the SLM was activated when the PSF was modulated, or deactivated, i.e. turned off, which recapitulates the standard PSF behavior seen for a conventional microscope, as the deactivated SLM acts as a mirror.

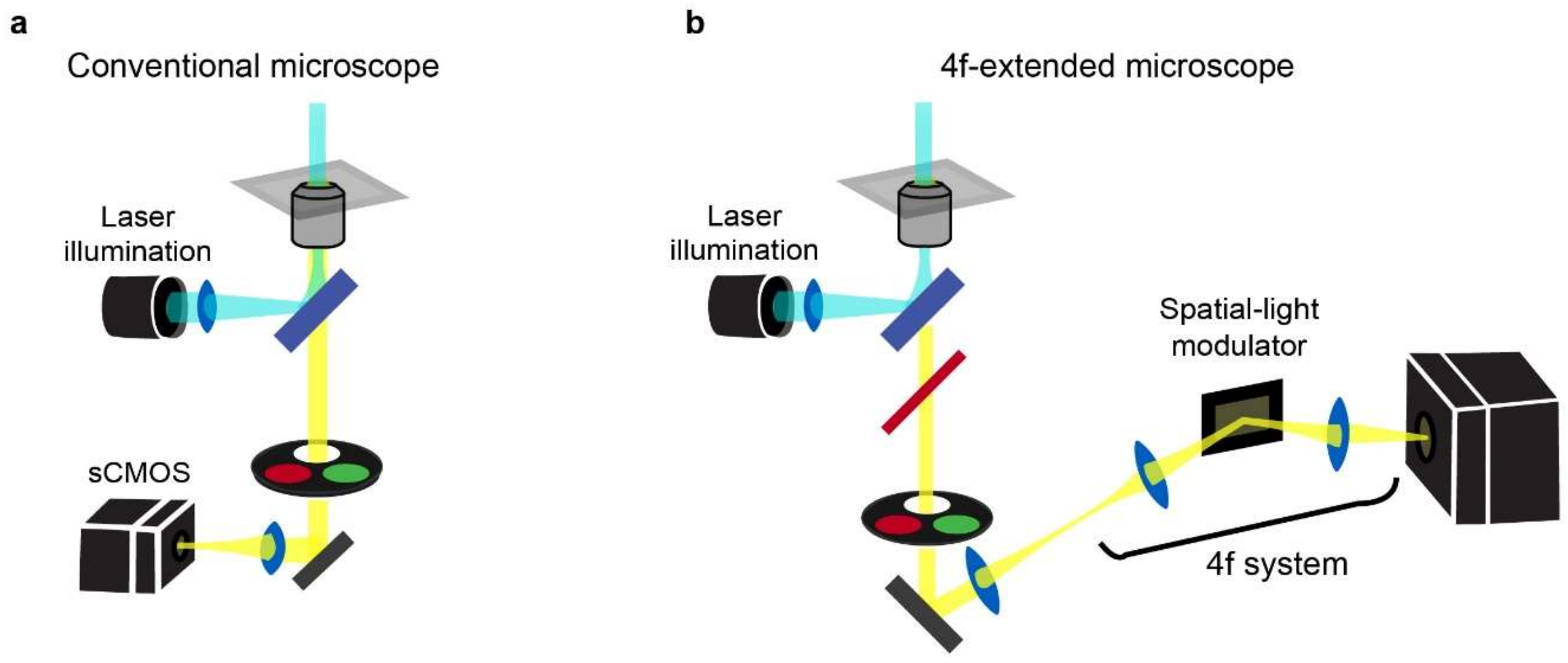

**Fig. S10** Microscope schematic for the **a** standard PSF, **b** optimized PSF.